\documentclass{article}

\usepackage[utf8]{inputenc} 
\usepackage[T1]{fontenc}    
\usepackage{lmodern}

\usepackage{lineno}   
\usepackage{amsmath}  
\usepackage{etoolbox} 

\newcommand*\linenomathpatch[1]{%
  \cspreto{#1}{\linenomath}%
  \cspreto{#1*}{\linenomath}%
  \csappto{end#1}{\endlinenomath}%
  \csappto{end#1*}{\endlinenomath}%
}

\linenomathpatch{equation}
\linenomathpatch{gather}
\linenomathpatch{multline}
\linenomathpatch{align}
\linenomathpatch{alignat}
\linenomathpatch{flalign}

\usepackage{authblk}
\usepackage{fullpage}
\usepackage{graphicx}

\usepackage{titlesec}

\titleformat{\section}[hang]{\Large\bfseries}{\thesection.}{1em}{}
\titleformat{\subsection}[hang]{\large}{\thesubsection.}{1em}{}
\titleformat{\subsubsection}{\bfseries\itshape}{\thesubsubsection.}{0em}{}

\usepackage{breakcites}
\usepackage[round]{natbib} 
\usepackage{placeins} 

\usepackage{amsmath}
\usepackage{amssymb}
\usepackage{amsthm}
\usepackage{tikz}

\usepackage{enumitem}

\usepackage{hyperref}       
\usepackage{url}            
\usepackage{svg}
\usepackage{booktabs}       
\usepackage{amsfonts}       
\usepackage{microtype}      
\usepackage{xcolor}         
\usepackage[export]{adjustbox}
\usepackage[mode=buildnew]{standalone}
\usepackage{caption}
\usepackage{subcaption}
\usepackage[bottom]{footmisc}
\usepackage{float}
\usepackage{titletoc}

\renewcommand{\cite}{\citep}

\hypersetup{
    colorlinks=true,
    linkcolor=cyan,
    citecolor=cyan,
    urlcolor=cyan,
}

\title{Dynamic Causal Models of Time-Varying Connectivity}
\author[1,*]{Johan Medrano}
\author[1]{Karl J. Friston}
\author[1]{Peter Zeidman}

\affil[1]{Functional Imaging Laboratory \protect\\ Department of Imaging Neuroscience, UCL Queen Square Institute of Neurology\protect\\ 12 Queen Square, London WC1N 3AR, UK.}
\affil[*]{Corresponding author}
\date{}

\begin{document}

\maketitle

\begin{abstract}
    This paper introduces a novel approach for modelling time-varying connectivity in neuroimaging data, focusing on the slow fluctuations in synaptic efficacy that mediate neuronal dynamics. Building on the framework of Dynamic Causal Modelling (DCM), we propose a method that incorporates temporal basis functions into neural models, allowing for the explicit representation of slow parameter changes. This approach balances expressivity and computational efficiency by modelling these fluctuations as a Gaussian process, offering a middle ground between existing methods that either strongly constrain or excessively relax parameter fluctuations. We validate the ensuing model through simulations and real data from an auditory roving oddball paradigm, demonstrating its potential to explain key aspects of brain dynamics. This work aims to equip researchers with a robust tool for investigating time-varying connectivity, particularly in the context of synaptic modulation and its role in both healthy and pathological brain function.

\end{abstract}

\section{Introduction}

    Advances in neuroimaging technology now enable the recording of longer experimental datasets, providing unprecedented opportunities to explore the functional significance of slow fluctuations in brain dynamics ~\cite{bassett2011dynamic,hutchison2013dynamic,zarghami2020dynamic,vidaurre2018discovering}. These fluctuations arise from various mechanisms unfolding over extended periods, such as shifts in extracellular ion concentrations or changes in synaptic efficacy that reflect adaptation and learning processes~\cite{jefferys1995nonsynaptic,abraham1996metaplasticity,citri2008synaptic,magee2020synaptic}. Modelling these slow processes has provided insight in the mechanisms of epilepsy~\cite{breakspear2006unifying, jirsa2014nature, rosch2018calcium, wendling2016computational,papadopoulou2015tracking} and anaesthesia~\cite{ching2014modeling, adam2023modulatory, soplata2023rapid}. Such slow dynamics are also relevant in the context of predictive coding, where alterations in synaptic efficacy are thought to signify the brain's adaptation to stimuli variability and the encoding of uncertainty or precision~\cite{auksztulewicz2016repetition, friston2012dopamine, moran2013free}. In addition, these slow endogenous fluctuations may also introduce confounding effects in neuroimaging data that could confound experimental effects if not modelled properly.

    Regardless of the recording modality, slow fluctuations  changes in neural activity and confounds  neuronal coupling will are expected to impact the measured signals in a nonlinear fashion and need to be acknowledged  assessed to ensure the soundness  validity of any statistical analysis. This can be addressed by constructing statistical models of how the  these data are generated, i.e., building \emph{generative models} that explicitly accomodate these slow fluctuations~\cite{medrano2024linking}. Far from the ambition of perfectly tackling all the complexities underlying brain dynamics, generative models can provide an idealised summary of the key mechanisms generating observed brain responses, recapitulating the activity in a sparse set of regions using a few parameters with a clear biophysical meaning, such as average number of synaptic projections or average membrane potentials of a population. These parameters can be refined in light of the data, allowing one to tailor the model to a particular experimental recording. This is the approach taken by Dynamic Causal Modelling (DCM), a popular modelling framework leveraging Bayesian statistics to test hypotheses about the mechanisms generating the experimental neuroimaging data at hand~\cite{friston2003dynamic, kiebel2008dynamic}.  
    The present work follows the generative modelling philosophy to account for slow changes in synaptic efficacy by explicitly introducing time-varying parameter fluctuations in dynamic causal models. 
    
    This problem is not new and has been tackled in different ways. For instance, Auksztulewicz and Friston~\cite{auksztulewicz2016repetition} were interested in the slow changes in self-inhibition and synaptic gain underlying adaptation in an auditory roving oddball paradigm. They augmented existing neural mass models used in DCM, by allowing connectivity parameters to be weighted by a linear mixture of hypothesised effects or covariates. These covariates were organised in a  \emph{design matrix}, with two columns, expressing hypothesised monotonic and a phasic changes in effective connectivity. By estimating the weights associated with each of these two distinct modulatory effects per connection, they uncovered  found that self-inhibition underwent phasic effects while extrinsic (between node) effective connectivity changed monotonically. In a different application, understanding drug-induced onset of epilepsy, Rosch et al.~\cite{rosch2018calcium} used a windowing approach to model time-frequency spectral power, fitting a DCM to each time bin and connecting time points using a random effects linear model (parametric empirical Bayes, PEB). This approach was later formalised by Jafarian et al.~\cite{jafarian2021adiabatic} in what is today known as adiabatic DCM. 

    While both approaches are of interest in specific situations, they suffer from some limitations. The design matrix approach of~\cite{auksztulewicz2016repetition}, as used in their work, has the drawback of needing to find  specifying an explicit set of basis functions, e.g., phasic and monotonic, whose shape is a-priori specified such that only their amplitude is evaluated estimated. This is perfectly suited when the slow changes in connectivity have a known shape or form, but not when their form is unknown -- in other words, this approach is highly constrained. On the other hand, adiabatic DCM does not explicitly model slow variations in connectivity in the first instance. Rather, it inverts a model per time window. The estimated parameters are then joined together over time using the linear random effects ``PEB'' model. By construction, this approach is more expressive, as parameters are allowed to vary on a per-window basis.  However, it may be too expressive, as the constraint that the parameters should evolve slowly and smoothly are only introduced post-hoc, at the between-window level, following the estimation of each individual window's model. In addition, this approach is computationally demanding, necessitating the estimation of a single model per window. Furthermore, recovery of window-specific parameters from the PEB model is based on a quadratic (i.e., Laplace) approximation, which can produce inconsistent results in the context of nonlinearities of certain ‘brittle’ DCMs.

    In this work, we aim at formalising a relatively simple approach which builds on the previous work introduced above and might seem quite natural to the experienced DCM user. We propose to construct a design matrix using an expressive set of temporal functions — up to a sufficient order —  to capture  model only the slow fluctuations in model  connectivity parameters. These fluctuations are captured by modulatory parameters (matrix $B$ in the DCM neural models), which weight temporal basis functions that capture  model the slow trajectory of a particular connection. Estimating these modulatory parameters uncovers  furnishes the \emph{spectral expansion} of the slow connectivity trajectory. As is common in DCM, these effects are normally distributed, henceforth  effectively modelling the time-varying connectivity as a Gaussian process. While this approach provides only the required level of expressivity, thereby minimising complexity, it still allows us to retain the stochastic aspect of the uncovered  underlying trajectory. Thus, the time-varying connectivity can subsequently be used with PEB, in the same spirit as  of adiabatic DCM. In addition, the number of temporal basis functions might be significantly lower than the number of time points, thus having far fewer free parameters than adiabatic DCM and therefore much better  greater statistical power  efficiency (via reduced model complexity). Finally, the proposed approach is constructed on mechanisms that exist in most variants of DCM, and might therefore be useful for most neuroimaging modalities using standard DCM implementations.  
    
    This paper is divided as follows. We first introduce Materials and Methods, starting by introducing neural mass models of cortical activity, then moving onto DCMs and their specific types of them, and finally presenting our approach for modelling time-varying connectivity. Then in  In the Results section, we provide numerical validation  validations of our approach, using both simulated evoked and spectral local field potential data. We then present some results on modelling empirical evoked responses in an auditory roving oddball paradigm. Finally, we discuss the main modelling decisions that come with this approach, such as selecting the basis set or its order.
We hope that this work will provide neuroimaging researchers with a new tool for investigating time-varying connectivity, and help to uncover the mechanisms underlying synaptic gain modulation and facilitation in both healthy and pathological subjects.

\section{Materials and Methods}
    \subsection{Neural Mass Models}
        Cortical neurons are organised in columns that extend across the different layers of the cortex. These columns are characterised by local (\textit{intrinsic}) connectivity patterns --- e.g., the average number of synaptic projections between distinct neuronal populations in different cortical layers --- that differ from global (\textit{extrinsic}) patterns of projections among different cortical regions. Within a cortical column, pyramidal neurons in superficial and deep layers contribute largely to the electromagnetic signals measured by LFP, EEG, and MEG~\cite{kiebel2008dynamic,david2005modelling}. Their activity is modulated by complex interactions with other populations within the columns, notably inhibitory interneurons and spiny stellate cells. Modelling electromagnetic signals originating from cortical sources necessitates accounting for these intrinsic interactions. 

        Neural Mass Models (NMMs) allow us to describe the electrical activity of a cortical column as a point source or  equivalent current dipole. For this, NMMs typically describe the average activities of a curated set of neuronal populations, and their evolution due to intrinsic interconnections and extrinsic inputs. NMMs come in two different strains, either describing the dynamics of the currents and membrane potentials -- \textit{convolution-based models} -- or the dynamics of the currents and conductances -- \textit{conductance-based models}~\cite{moran2013neural}. While convolution-based models are typically simpler and  more computationally expedient, conductance-based models allow for a more detailed description of the neurotransmitter receptors dynamics. Ultimately, both types of NMMs allow us to model the complex dynamics of the average membrane potential of pyramidal neurons, which can then be used to model LFP and M/EEG signals. 

        NMMs model the activity of cortical columns from either three or four neuronal populations. Three-population models comprise excitatory pyramidal neurons that are reciprocally connected with excitatory interneurons (e.g., spiny stellate cells in granular layer IV) and inhibitory interneurons (pooled across layers)~\cite{jansen1995electroencephalogram,david2005modelling,moran2013neural}. In four-populations NMMs, distinct variables are used to describe the activity of superficial (layer I/II) and deep (layer V/VI) pyramidal neurons~\cite{bastos2012canonical}. In these canonical microcircuit (CMC) models, the separation of superficial and deep pyramidal neurons gives additional flexibility, allowing them to distinguish stronger and faster superficial contributions to the local electromagnetic field from the slower and weaker contributions of deep pyramidal neurons. Biological CMCs are assumed to play a key role in cortical computation, serving as building blocks of hierarchical predictive coding in the cortex~\cite{bastos2012canonical, friston2008hierarchical}.  
        
    \subsection{Dynamic Causal Models}      
        Dynamic Causal Modelling (DCM) is a framework for estimating the parameters of a system of coupled differential equations -- such as neural mass models -- from empirical data using (variational) Bayesian statistics. In neuroimaging, DCM is used to recover the directed influence -- or \textit{effective connectivity} -- between different brain regions. In DCM, the difference in brain responses between experimental conditions is modelled as resulting from condition-specific modulation of the effective connectivity. Bayesian model inversion allows one to estimate the extrinsic (between-regions) and intrinsic (within-region) connectivity and their modulation by experimental conditions, alongside other model parameters. 

        \subsubsection{Effective connectivity and modulatory effects.}
        DCM captures interconnected current sources giving rise to M/EEG and LFP data using networks of NMMs. These networks may feature different types of connections, each reflecting standard patterns of extrinsic (between-region) connectivity found between cortical neuronal populations at different locations. Extrinsic projections are either forward (\textit{ascending}), backward (\textit{descending}), or lateral, and describe connections towards, respectively, increasing, decreasing, or a similar level of the cortical processing hierarchy. 
        
        In DCM, the efficacy of synaptic projections from a source population in a source region to a target population in a target region is defined by a constant --- e.g., $E_0$ --- which is scaled (multiplied) by a log-scaling factor --- e.g., $E$ --- that is estimated from the data. Given $E$, the resulting synaptic efficacy is $E_0\exp(E)$. This transformation ensures that the synaptic efficacy can be tuned up or down -- by, respectively, a positive or negative log-scaling factor $E$ -- without changing its sign. DCM uses this log-scaling of connectivity parameters to estimate baseline connectivity and between-condition modulatory effects with a simple linear model. The (log-scaling) efficacy of synaptic projections $E_{ck}$ for connection type $c$ and condition $k$ is given by
        \begin{align}
            \label{eq:effective-connectivity}
            E_{ck} = A_c + \sum_{m=1}^M B_m X_{km} 
        \end{align}
        For a model of $n$ sources, $E_{ck}$, $A_c$, and $B_m$ are all $n$ by $n$ matrices whose $(i,j)$-element captures an aspect of the connection from $j$ to $i$. For instance, $E_{ck}$ is a matrix whose $(i, j)$ element represents the efficacy of connections of type $c$ from region $j$ to region $i$ in the condition $k$. $A_c$ represents the log-scaling factor of the baseline efficacy of type-$c$ connections. The summation term collects all of the modulatory effects appearing in condition $k$. This term takes a positive value when the synaptic connectivity increases in condition $k$ as compared its baseline value, and takes a negative a value when the connectivity decreases from baseline. The overall modulatory  effect in condition $k$ is expressed as a linear combination of $M$ distinct modulatory effects, each stored in one of the $B_m$ matrices, and dispatched to the condition $k$ by 
        the \textit{design matrix} $X$. The design matrix can be used to constrain the parameterisation of the model, e.g., to estimate a single modulatory effect across several conditions. 

        \subsubsection{Model estimation}
        A typical DCM is constructed by selecting the type of NMM, the locations of the different sources, and the type of connections between them. Then, one uses the design matrix $X$ and modulatory effect matrices $B_m$ to specify which connections undergo condition-specific modulations in which condition. From this description, DCM constructs prior Gaussian distributions $p(\theta)$ for all parameters $\theta$ in the model. DCM then evaluates the model at the current mean value of each parameter to construct predictions about the data. A precision (inverse variance) parameter controls the tolerance for the spread of the observed data around the model predictions, giving the predicted data distribution or likelihood $p(y|\theta)$. From this, the current parameter distribution is scored by a free-energy functional. This free-energy provides a lower bound on the intractable \textit{model evidence} (a.k.a., marginal likelihood) -- i.e., how well the model is able to explain the data: 
        \begin{align}
            \underbrace{\ln p(y)}_{\text{model evidence}} \geq F(y, m) = \underbrace{\mathbb{E}_{\theta \sim q}\left[\ln p(y|\theta) \right]}_{\rm accuracy} - \underbrace{D_{\rm KL}\left(q(\theta)|p(\theta)\right)}_{\rm complexity}
        \end{align}
        The free-energy reflects the trade-off between accuracy and complexity, favouring simple models of the data. The mean and variance of the \textit{approximate posterior parameter distribution} $q(\theta)$ are then updated towards the maximum of the free-energy. By repeating model evaluation and updates, DCM iteratively constructs a Gaussian approximation to the \textit{posterior} distribution of parameters, i.e., their distribution after observing the data. This \textit{variational inference} procedure provides a tractable approximation to the intractable \textit{Bayesian inference} problem. In addition to an estimate of the posterior parameter distribution, variational inference yields the maximum free-energy, which approximates the model evidence $\ln p(y)$ and can be used to compare models --- a more positive free-energy for the same data reflecting a better model. 

        \subsubsection{Models of evoked responses.} 
        Evoked responses refer to patterns of LFP and M/EEG responses that are time-locked to the onset of an external stimuli. These responses are generated by the synchronous firing of thousands of neurons, which can be locally described using NMMs. Evoked responses typically propagate through the cortical hierarchy, highlighting the hierarchical predictive processing of the stimulus. This makes evoked responses an important tool for studying predictive coding and sensory processing in the brain. In particular, the patterns of propagation and the overall shape of the evoked response at different level of the cortical hierarchy are subtle indicators of the directed coupling between the different regions. By using explicit models of the activity in each region -- i.e., NMMs -- and asymmetric extrinsic coupling between cortical regions, DCM has been successfully used to infer the directionality of the coupling between different regions from evoked responses --- and consequently, the relative position of these regions within the cortical hierarchy~\cite{brown2013functional,bastos2015dcm,garrido2008functional}. For this, DCM integrates the response of the network of NMMs to a Gaussian bump function, which models stimulus onset. This bump generates evoked responses over the scalp. DCM explains between-condition variability in the response by changes in the synaptic efficacy both intrinsically (through self-inhibition of the populations) and extrinsically (through cross-regional connections). 
            
        \subsubsection{Models of spectral responses.}
        Models of spectral responses in M/EEG and LFP have been proposed to capture the complex patterns of frequency spectra and spectral coherence observed over the scalp~\cite{friston2012dcm, moran2011consistent, moran2009dynamic}. In this case, spectral features can help disambiguate between the activity of deep and superficial pyramidal neurons, also helping with identifying the right types of connections between two regions. In addition, the phase information of cross-spectral density is a good indicator of the propagation delay between regions. 
        
        In DCM of spectral responses, regions of interest are viewed as filtering background activity with a scale-free power law~\cite{friston2012dcm}. The overall response is tuned by the intrinsic and extrinsic connectivity. To compute the spectrum, the differential equations of the coupled NMMs are linearised around the fixed point determined by the current values of parameters. This linearisation allows the transfer function transforming the noise into the observed spectrum to be explicitly derived. Similarly to other DCMs, between condition variability in spectral responses is accounted for by changes in effective connectivity.  
        
    \subsection{Time-Varying Connectivity}
        \label{sec:time-varying-connectivity}
        Over the course of long neuroimaging recordings, the synaptic efficacy between different neuronal populations is likely to change. These changes might be uncontrolled, for instance due to fatigue or dehydration, or reflect cognitive processes of interest, such as learning and attention. Some of the slow changes, such as the variation of extracellular potassium, can develop smoothly before triggering abrupt changes in brain dynamics, e.g., epileptic seizures~\cite{jirsa2014nature, wendling2016computational, rosch2018calcium}. All these conditions can lead to complex multiscale phenomena due to circular causality --- for instance, increased neuronal activity might accelerate the accumulation of extracellular potassium, in turn influencing the neuronal activity. This motivates developing simple models based on NMMs while accounting for time-varying effective connectivity.  

        \subsubsection{Empirical motivations}
        To construct NMMs with time-varying connectivity, we first look at the necessary requirements for a separation of temporal scales. Indeed, 
        the phenomena described above are most conveniently studied under a separation of temporal scales, i.e., by separating slow itinerant variables --- that dictate effective connectivity --- from fast convergent ones --- that describe neuronal activity~\cite{haken1977synergetics,haken2006synergetics}. Under such a decomposition, slow variables completely determine the dynamics of the fast variables~\cite{medrano2024linking}
        . The real part of Lyapunov exponents -- the values that determine the rate of convergence to stability of a dynamical system --  of the NMMs need to be at least an order of magnitude more negative than that of the synaptic connectivity dynamics~\cite{carr2012applications}. NMMs therefore need to exhibit fast-enough activity, with linear rates in the order of 1 to 10ms, allowing for separating fast changes in current, membrane potentials, or conductance, from the "slow" changes of synaptic efficacy that unfold over few 100 ms or more. 
        
        While we have committed to NMMs to describe fast dynamics, we need to consider how to describe the slow changes in synaptic efficacy. These could be described using a dynamical system. However, this might make the model too complex to be accurately estimated from empirical data, as the tiniest shift in the synaptic efficacy dynamics would result in large changes in the modelled dynamics. Due to their strong nonlinearities, DCMs based on NMMs already suffer from local minima problems that can hinder convergence and make them challenging to estimate from data. To avoid adding additional complexity --- and the associated convergence problems --- to these models, we propose to use sets of slow temporal basis functions --- in a linear  model --- to capture time-varying connectivity. This approach can be implemented using the existing modulatory effect mechanism in DCM, thereby eschewing any significant modification of the existing DCM code or its formulation. 
        
        \subsubsection{Technical details}
        The proposed approach only requires changing what is considered to be a "condition" in DCM. As mentioned previously, DCM allows condition-specific effects to be modelled using modulatory $B$ matrices that are mapped onto conditions by a design matrix $X$. The rationale of our approach is simple: if we epoch data in consecutive time bins and consider each bin as a "condition", then the design matrix maps each modulatory effect onto different time points, thus modelling modulation of the connectivity \textit{over time}. By selecting adequate forms of design matrix, we can express time courses of modulatory effects that evolve over time. Moreover, by projecting the (normal) distribution of the $B$ matrix components onto the temporal basis functions, one can retrieve the (normal) distribution of the time-course of connectivity.

        Our approach simply gives another reading to Eq.~\eqref{eq:effective-connectivity}, giving the effective connectivity for connections of type-$c$ at time $t$ as: 
        \begin{align}
            E_{c}(t) = A_c + \sum_{m=1}^M B_m X_{m}(t) 
        \end{align}
        In practice, $A_c$ and $B_m$ are  normally-distributed matrix-valued random variables. Thus, the trajectory of effective connectivity is a multivariate Gaussian process, with mean   
        \begin{align}
            \mathbb{E}\left\{E_{c}(t)\right\} = \mathbb{E}\left\{A_c\right\} + \sum_{m=1}^M \mathbb{E}\left\{B_m\right\} X_{m}(t) 
        \end{align}
        and covariance
        \begin{align}
            \text{Cov}\left\{E_{c}(t), E_{c}(s)\right\} = \text{Cov}\left\{A_c, A_c\right\}  + \sum_{m=1}^M \sum_{n=1}^M  \text{Cov}\left\{B_m, B_n\right\} X_{m}(t) X_{n}(s) 
        \end{align}
        where the covariance between matrices is a 4-tensor whose $(i,j,k,l)$ element is given by: 
        \begin{align}
            (\text{Cov}\left\{P, Q\right\})_{ijkl} = \mathbb{E}\left\{\left(P_{ij} - \mathbb{E}\left\{P_{ij}\right\}\right)\left(Q_{kl} - \mathbb{E}\left\{Q_{kl}\right\}\right)\right\}
        \end{align}  
        This approach effectively allows for Bayesian estimation of a matrix-valued Gaussian process capturing the time-varying connectivity. The temporal basis functions that compose the design matrix determine the type of trajectory that can be expressed, as well as the covariance structure over time.

        \subsubsection{Analytical example}
         Let us assume a time series of neuronal activity of duration $T$ is divided into $K$ epochs, each with mid-time $t_k$, and possibly different durations. To model time-varying effective connectivity, our approach requires us to first select an appropriate temporal basis set. A simple choice of basis set is the Fourier set. We can construct a design matrix with a Fourier set of a given (even) order $M$ by stacking columns of sines and cosines of frequency 1 to $M/2$: 
        \begin{align}
            X_{m}(t) = \begin{cases}
                \cos(2\pi m t/ T) & \text{ if } m \text{ odd}, \\
                \sin(2\pi m t/ T) & \text{ otherwise. }
            \end{cases}
        \end{align}
        Each row of the design matrix is constructed by evaluating the basis set at the mid-time $t_k$ of each epoch, giving a matrix with $K$ rows and $M$ columns, i.e., $X_{km} = X_{m}(t_k)$. Under such a design matrix, the model of time-varying effective connectivity becomes: 
        \begin{align}
            E_c(t_k) = A_c + \sum_{m \text{ odd}} B_{m} cos(2 \pi m t_k / T) +  \sum_{m \text{ even}} B_{m} sin(2 \pi m t_k / T)
        \end{align} 
        We see that using the Fourier set as the design matrix expresses the Fourier series expansion of the time-varying connectivity, with the modulatory effect matrices capturing the real part, for odd $m$, and the imaginary part, for even $m$, of the Fourier coefficient. Interestingly, the order of the basis set determines the maximal frequency of the trajectory, and makes explicit the choice of the temporal scale over which we allow effective connectivity to fluctuate -- assuming that its spectral power is contained over the slow frequencies. Moreover, it is worth noting that we explicitly evaluate the basis set at each time point, akin to a Lomb-Scargle (least-squares) estimator of the periodogram~\cite{lomb1976least,scargle1982studies}. This removes the need for an even sampling interval, and thus does not require one to construct epochs of the same duration~\cite{vanderplas2018understanding}. In practice, one usually replaces the Fourier basis functions with the corresponding discrete cosine set, due to its superior compression characteristics (i.e., replacing the cosine and sine of one frequency with cosines at two frequencies).

\section{Results}
    \subsection{Numerical validation}
        The previous section proposed that time-varying effective connectivity could be aptly modelled using a spectral expansion on a set of temporal basis functions, leveraging existing modulatory effects in DCM. In this section, we use simulated data to establish the face validity of our approach; namely, that we can recover time-varying connectivity used to simulate the data. 
        
    \subsubsection{Simulation setup}
        To evaluate our approach, we constructed a model of two connected cortical regions. Each region consists of a CMC model -- a four-population NMM separating deep and superficial pyramidal neurons. The first region has forward connections to the second region, and receives backward projections from it. Forward connections are characterised by abundant synaptic projections from superficial pyramidal neurons to spiny stellate cells in middle layers of the second region. Backward connections have strong connections from deep pyramidal neurons of the second region to the superficial pyramidal and inhibitory interneurons in the first region. This reciprocal connectivity structure is typically observed between a lower sensory region -- from which forward connections originate, here region 1 -- and a higher cognitive one.
        
        We assume that the first region receives sensory inputs through thalamic projections targeting the spiny stellate cells in the middle layers. This input excites region 1 and propagates through forward projections to region 2. The neuronal activity in both regions is tracked through local field potentials, assumed to be measured from the surface of the scalp by electrodes LFP 1 and LFP 2. Similarly to previous work, the LFP signals are assumed to result from the average membrane potential of pyramidal neurons in superficial layers, plus some additive white Gaussian noise. 

        We validate our approach on both simulated and spectral responses. For simulating both types of responses, we introduce slow fluctuations of the effective connectivity that last over the entire time course of each simulation. These fluctuations were introduced using the modulatory effect mechanisms in DCM, as described above. In these simulations, we focus on modulating both the self-inhibition of superficial pyramidal neurons and the projections from superficial pyramidal neurons to spiny stellate cells.

        To recover the time-varying connectivity, we specified a model with a sixth-order cosine set as the design matrix. Element $(k,m)$ of the basis set, corresponding to time point $t_k$ and basis function $m$, is given by $\cos(t_k m \pi / T)$. The contribution of the $k$-th basis set in explaining the fluctuation of each connection is captured by the $m$-th modulatory effect. We run the standard DCM inversion scheme to fit the model to the data and obtain an approximate posterior estimate of modulatory effect matrices and other model parameters. By projecting the mean and covariance of each modulatory effect matrix onto the basis set, we retrieve the mean and covariance of the time course of effective connectivity.

    \subsubsection{Evoked responses}
        We first generate and recover evoked responses with time-varying connectivity. This setup simulates experimental conditions in which an experimental factor, such as stimulus repetition, induces a change of synaptic efficacy that is reflected by a progressive change in the shape of the subsequent evoked responses. This kind of model may be useful for studying stimulus habituation and learning.  
        \begin{figure}[t]
            \centering
            \includegraphics[width=\linewidth]{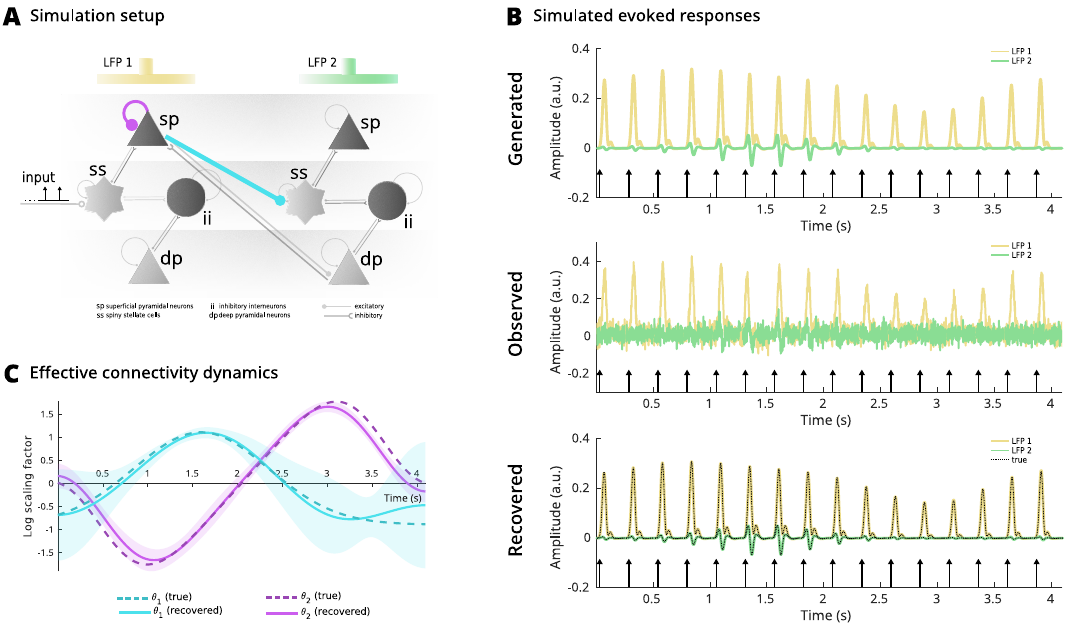}
            \caption{Face validity  of time-varying connectivity model for evoked responses. A. Simulation setup consisting of two interconnected regions, generating signals at channels LFP 1 (yellow) and LFP 2 (green). The stimulus enters in spiny stellate cells of region 1. Slowly fluctuating connections are self-inhibition of superficial pyramidal neurons (purple) and forward connections to spiny stellate cells (blue). B. Evoked responses simulation and recovery. The first row displays the generated (noise-free) ERPs at LFP 1 (yellow) and LFP 2 (green). Stimulus onset are displayed with a black arrow. The second row shows observed ERPs at both channels. The third row shows ERPs predicted by the model, with the black dotted lines representing the generated (``true'') ERPs. C. Time course of the effective connectivity. The blue and purple dashed lines show the time course of connectivity used to generate the data, with line colours corresponding to the connections highlighted in panel A. The corresponding solid lines represent the recovered mean of the connectivity trajectory, with the shaded areas showing the posterior standard deviation. }
            \label{fig:erp-simulated}
        \end{figure}
        
        We generated evoked responses by simulating 16 stimuli targeting region 1. Each stimulus is modelled by a Gaussian bump function centred at 64ms after each stimulus onset with a 32ms standard deviation, and an interstimulus interval of 500ms. The time course of connectivity modulation is shown in Figure~1.C. The model was integrated to generate the (noiseless) time course of each population's activity between 0 and 256ms after stimulus onset. We added white Gaussian noise to the average membrane potential of superficial pyramidal neurons to generate the "observed" LFP signal. We then used these data to invert the DCM with temporal basis functions and project the modulatory effects on the design matrix to construct the time-varying  connectivity distribution. The results are presented in Fig.~\ref{fig:erp-simulated}.

        In Fig.~\ref{fig:erp-simulated}.B., we notice that the model fit is particularly good, as DCM manages to recover the generated ERPs from the noisy observed LFP signals almost perfectly. In particular, the estimated model manages to capture the modulation of the ERP shape, with an increase in the response amplitude of both regions at around 1.5 seconds and an attenuation of the response amplitude in both regions at around 3 seconds. The recovered time-varying connectivity is displayed on  Fig.~\ref{fig:erp-simulated}.C. We see that the distribution of the recovered time-varying connectivity matches the true trajectory. This means that the model correctly captured the relative strength of modulatory effects in both regions that is necessary to explain the two observed noisy LFP signals. We also note that the model   to recover the first and last 0.5 seconds for the self-inhibition of superficial pyramidal neurons in region 1.  
    
    \subsubsection{Spectral responses}
\begin{figure}[t]
            \centering
            \includegraphics[width= \linewidth]{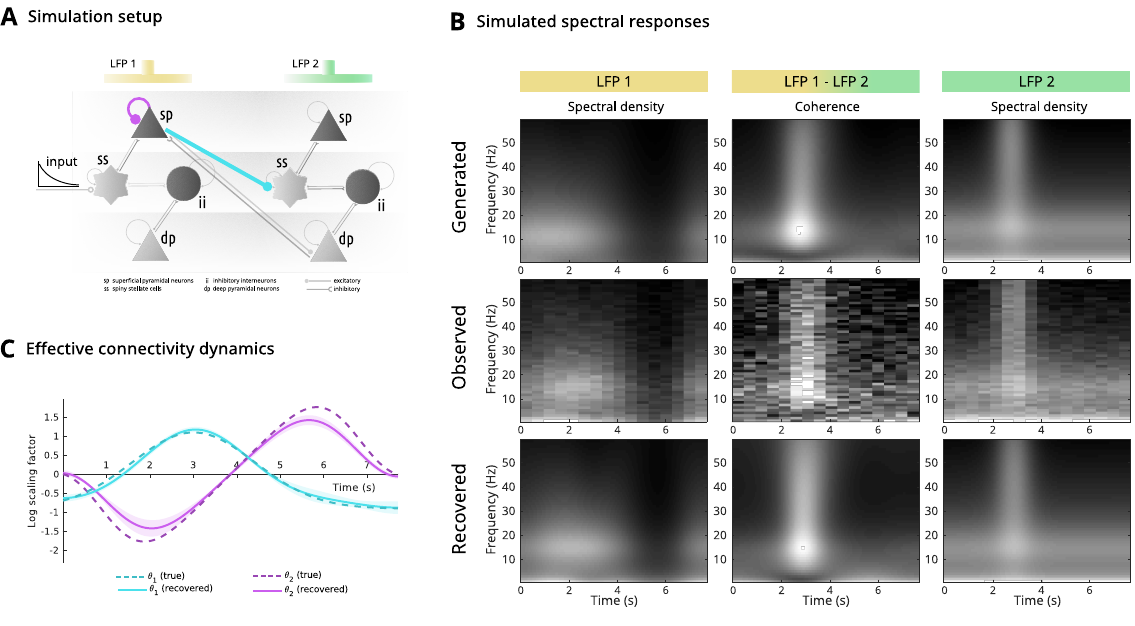}
            \caption{Face validity  of time-varying connectivity model for spectral responses. A. Simulation setup consisting of two interconnected regions, generating signals at channels LFP 1 ( yellow) and LFP 2 (green). The background scale-free noise enters in spiny stellate cells of both region. Fluctuating connections are self-inhibition of superficial pyramidal neurons (purple) and forward connections to spiny stellate cells (blue). B. Spectral responses simulation and recovery. The first row displays the generated (noise-free) time-frequency image of spectral density at LFP 1 (yellow) and LFP 2 (green), and the time-frequency image of the cross-spectral density.  The second row shows observed responses. The third row shows spectral responses and coherence recovered by the model. C. Time course of the effective connectivity. The blue and purple dashed lines show the time course of connectivity used to generate the data, with line colours corresponding to the connections highlighted in panel A. The corresponding plain lines represent the recovered mean of the connectivity trajectory, with the shaded areas showing the posterior standard deviation.  }
            \label{fig:tfm-simulated}
        \end{figure}

        Next we generate the cross-spectral densities at each LFP channel under time-varying connectivity. This simulation can be thought of as modelling progressive changes in synaptic connectivity induced by experimental changes, or following natural itinerance as in resting state or naturalistic experiments. For this, we specify background noise input to each region. The noise is assumed to have a $1/f$ power spectral density, typical of scale-free background brain activity. Similarly to our evoked responses simulation setup, we introduce slow fluctuations of connectivity over the entire simulation time. The simulation time is set to 7.5  seconds and we epoch the data into 16 time bins. We add chi-squared noise with a power law amplitude to the cross-spectral densities, with random variables drawn independently over frequency and time. We use this data to invert the DCM with temporal basis functions and project the posterior distribution of modulatory effects onto the basis set to recover the effective connectivity dynamics. Results are presented in Fig.~\ref{fig:tfm-simulated}.

        In Fig.~\ref{fig:tfm-simulated}.B., we observe that the spectral density at both regions is well recovered from the observed spectrum -- with some minor differences. Naturally, these differences combine and amplify when computing the cross-spectral coherence, which reflects the spectral coupling between the two regions. In Fig.~\ref{fig:tfm-simulated}.C, we see that the recovered trajectory for the effective connectivity is relatively consistent with that used to generate the data. The recovered trajectory seems less smooth than for evoked responses, potentially due to time binning effects. As for evoked responses, the self-inhibition of superficial pyramidal neurons in region 1 is poorly recovered in the first and last second of the data.
        
\subsection{Application}
    The previous section has established the face validity of the proposed approach for both spectral and evoked responses DCMs. We now apply the proposed approach to model the time-varying connectivity between auditory brain regions in an auditory oddball experiment. We chose a particular paradigm called the roving paradigm to foreground the importance of modelling slow changes in effective connectivity. In roving paradigms, a continuous stimulus changes sporadically to elicit an oddball response. The same stimulus is then presented repeatedly until it becomes a new standard. The transition from an oddball to a standard is thought to be mediated by short-term changes in synaptic efficacy that underwrite perceptual inference and sensory learning. It is these slow (trial to trial) changes in effective connectivity the current methodology is designed to quantify. In this example, the fluctuations in synaptic efficacy were introduced by careful experimental design — and provide a potentially useful non-invasive assay of short-term plasticity in the brain.
    
    \subsubsection{Data description and preparation}
    We use magnetoencephalography (MEG) data collected with optically-pumped magnetometers (OPMs) at the Functional Imaging Laboratory and published online by Mellor et al.~\cite{mellor2023real}. Two subjects underwent an auditory roving oddball paradigm, consisting of 80 deviant tones for each of the four experimental blocks and exponentially-drawn sequence lengths. In this work, we analysed the data from the first two blocks of subject 1. The experimental paradigm is as follows: a short auditory tone (70ms) is repeated with a 0.5 second interval a certain number of times ("standards"), and randomly switches to another tone ("deviants"), eliciting error-related potentials in auditory-related regions (Fig.~\ref{fig:experiment}.A). The "deviant" tone becomes the new "standard", and the procedure is repeated until the desired total number of "deviants" is achieved. Data was collected using 17 dual-axis OPM-MEG channels placed at a fixed distance from the scalp. Sensors are held in place by a 3D-printed headcast constructed from the subject's anatomical magnetic resonance imaging (MRI), removing the need for coregistration. The data were sampled at 1kHz, and processed following Seymour et al.~\cite{seymour2021using}.  

    \subsubsection{Model specification}
    We use the model structure from Garrido et al. \cite{garrido2008functional} for the evoked brain responses during an auditory roving oddball task. This model features five regions: left and right auditory cortex (A1, left MNI: $[-42; -22; 7]$, right: $[46; -14; 8]$), left and right superior temporal gyrus (STG, left: $[-61; -32; 8]$, right: $[59; -25; 8]$), and right inferior frontal gyrus (IFG: [46; 20; 8]). The activity of each region is modelled as an equivalent current dipole using a canonical microcircuit model. The membrane potential of superficial pyramidal neurons was assumed to be the sole contributor to the measured source signal. Between-regions connectivity was set following Garrido et al.\cite{garrido2008functional}, and composed of within-hemisphere forward connections between A1 and STG and between STG and IFG, reciprocal backward connections, and between-hemisphere lateral connections. The driving inputs were assumed to target left and right A1. The modulatory effects were assumed to impact the self-inhibition of superficial pyramidal neurons in all five regions, together with all forward and backward connections. The resulting model is shown in Fig.~\ref{fig:experiment}.B. We model the evoked responses over a peristimulus window of 500ms, starting 50ms before stimulus onset. As for our previous simulations, we used a 5th-order cosine set to capture the time-varying connectivity. This allows us to capture fluctuations between 0.11Hz (0.5 cycles over a total period of 4.5 seconds) and 0.55Hz (2.5 cycles over 4.5 seconds). We use the last standard of every sequence as a baseline (i.e., the "0" of the log scaling factor), defining the basis set between the oddball and the second-to-last stimulus of every sequence. 

    \subsubsection{Results}
    \begin{figure}[t]
        \centering
        \includegraphics[width=\linewidth]{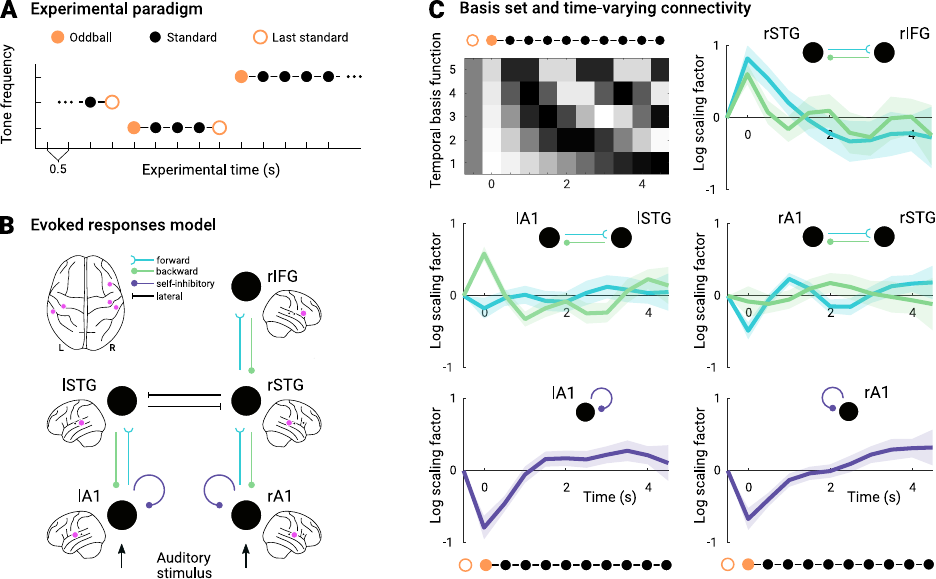}
        \caption{Application of DCM of time-varying connectivity to OPM-MEG data from an auditory roving oddball experiment. A. Illustration of the auditory roving oddball paradigm. Sequences of short auditory tones with 0.5 second interval  were presented to the participant. The first tone after switching tone frequency is an  oddball evoking a mismatch negativity response. B. Connectivity structure of the MMN model, inspired from Garrido~et~al.\cite{garrido2008functional}. The model contains left (l) and right (r) primary auditory cortex (A1), left and right superior temporal gyrus (STG) and right inferior frontal gyrus.   Coloured connections are modelled as modulated over time. Self-inhibition of STG and IFG were assumed as static (following \cite{garrido2008functional}) and were omitted on the diagram. C. Set of temporal basis functions and estimated time-varying connectivity. The upper left panel shows the design matrix filled with discrete cosine basis functions evaluated at each stimulus time after oddball onset, with the last standard being used as baseline. The five other panels shows the time course of the log scaling factor of  forward (blue), backward (green), and self-inhibitory (purple) connections, with shaded area indicating posterior standard deviation. Time is indexed from the oddball. }
        \label{fig:experiment}
    \end{figure}
    Results are presented in Fig.~\ref{fig:experiment}. We observe a negative gain on the self-inhibitory connections in primary auditory cortex (lA1 and rA1) at the onset of the oddball, returning to a small positive value over the course of the sequence. The forward connection between right STG and IFG follows the opposite trend. All remaining connections weakly fluctuate at frequencies below 1Hz, but the interval between stimuli is too large to analyse these fluctuations -- following the Nyquist criterion.  
    The decrease of self-inhibitory gain can be understood as explaining the excitation of primary auditory cortex when the oddball is presented. In a predictive coding account of oddball responses, this indicates a strong mismatch between the current predictions and observations, which is slowly resolved — on repeated exposure to a predictable stimulus — via experience-dependent plasticity.
    
\section{Discussion and Conclusion}
    This paper presents a new framework for modelling time-varying connectivity in DCM. Our approach rests upon a linear model for modulatory effects on effective connectivity that is built into DCM. By specifying temporal basis functions in a design matrix, we force modulatory effects to capture the spectral expansion of the time-varying connectivity. This improves on standard DCM analysis, by relaxing the unrealistic assumption that the strength of synaptic connections (effective connectivity) remains constant during a neural recording. Our approach improves on previous attempts at modelling time-varying connectivity, by simultaneously estimating parameters governing slow changes over seconds or minutes, and those governing fast changes in voltages and conductances over milliseconds. It does this using minimal assumptions that constrain only the timescale of the slow fluctuations. We expect this to be a useful modelling tool for a range of cognitive and clinical neuroscience applications.

    \subsection{Design choices}
    \subsubsection{Basis set selection}
    In practice, the choice of basis set and its order can be far from trivial. Conveniently, DCM provides an estimate of the log model evidence, called the free energy, which can be used to compare models. More precisely, the difference in free energy between two models is an estimate of the log-Bayes factor, quantifying the evidence towards one or the other model. Bayes factors are what most closely resemble a $p$-value in  Bayesian statistics, with a positive log-Bayes factor between model 1 and model 2 quantifying evidence towards model 1. This gives a straightforward and rigorous criterion to answer design choices, e.g., selecting a set of temporal basis functions and an order: a better model has a higher free energy after optimisation. Log-Bayes factors computed from differences in free energy form a rigorous statistical test -- indeed, it is guaranteed to be always equivalent to the most powerful test across all levels of sensitivity~\cite{neyman1933ix,fowlie2023neyman}. Notably, it can be used to ask whether connectivity is modulated over time. For this, it is sufficient to compare a model with an expressive basis set (for instance, the best model across all types and order of basis set), to a model without connectivity fluctuations. 

    \subsubsection{Condition-specific effects.}
    So far, we have solely considered applications in which one is either interested in modelling temporal fluctuations (with our approach) or different conditions (with a classical DCM). In practice, one might have several conditions and be interested in modelling, for instance, the difference in the trajectory of effective connectivity between conditions. We imagine that this type of model might be of interest in uncovering the trajectory of synaptic gain during learning two different stimuli, for instance differing in their variability. In that case, one can imagine constructing a "normal" design matrix for contrasting two conditions --- for instance, a two by two design matrix with one column expressing commonalities and the other expressing differences. By taking the Kronecker product between that matrix and the matrix containing the temporal basis set, one can construct a design matrix that would directly capture these between-condition differences over time. Indeed, this philosophy is similar to typical between-groups modelling with the general linear model, and should work robustly in DCM.   

    \subsubsection{Random effects and hierarchical models.}
    In some cases, one might want to first characterise the trajectory of time-varying connectivity, and then construct a generative model of it. Indeed, due to the decoupled structure of the free energy, previous work has shown that the contribution of the free energy to a particular level within a hierarchy of models depends only on the level below~\cite{friston2016bayesian}. This is the basis for parametric empirical Bayes, which allows one to estimate a linear model over the posterior parameter distribution of the model below~\cite{zeidman2019guide}. We have explained how our approach allows one to estimate the posterior distribution of time-varying connectivity as a Gaussian process. By construction, this Gaussian process can be used as stochastic observation in subsequent models. We anticipate that this could provide the grounds for constructing more realistic biophysical models of synaptic dynamics, using a hierarchical approach for computational expediency.

    \subsection{Perspectives}
    \subsubsection{Computational efficacy}
        Our approach reduces the computational complexity of estimating models of time-varying connectivity. Indeed, the number of basis functions is expected to be much lower than the number of modelled time bins. Therefore, the number of parameters per modulated connection is much lower than for, e.g., adiabatic DCM. Moreover, the number of time bins used can be optimised to match the span of the design matrix, i.e. such that the design matrix is square, giving Nyquist criterion in the case of Fourier set.  Then, by parallelising the computation of each time bin, we found the computational cost of inverting the model to be close to that of inverting a model without modulatory effects, as long as the number of rows in the design matrix (i.e., the number of columns if following the previous advice) is less than the number of processor cores. In practice, this optimisation --- made available in the latest SPM release --- makes our approach computationally appealing. 

    \subsubsection{Extension to fMRI}
        Because the proposed method is built directly on the modulatory effect mechanisms in DCM, one can imagine constructing models of fMRI timeseries with time-varying connectivity. We anticipate that this will be appealing, especially if combined with spectral DCMs for fMRI. This could enable analysing the dynamics of effective connectivity during resting state, but also during long experimental studies such as drug studies, and in pathological studies. It should nonetheless be kept in mind that the duration and delay of the  haemodynamic response generally smooths out the observed time series, such that the dynamics unfold over longer timescales. Therefore, although the same arguments about the separation of temporal scales can be brought forward, one should keep in mind that "slow" fluctuations in the case of fMRI might be much slower than for EEG and MEG signals. Aside from that, there are no obstacles to extending the current approach for time-varying connectivity to fMRI.

\subsection {Conclusion}
    This paper introduces and analyses a new approach to model time-varying connectivity. The approach leverages existing mechanisms for modelling modulatory effects in DCM. By using slow temporal basis functions to construct the design matrix of the DCM, we show that it is straightforward to recover the spectral expansion of the effective connectivity trajectory. We validate our approach on both simulated and real data, using DCMs of both evoked and spectral responses. We discuss possibilities for integrating our approach within more advanced statistical analysis pipelines, to select the basis set and its order, integrate condition-specific models of time-varying connectivity, and build more complex models of its dynamics. Our method is computationally efficient, accessible in SPM and other implementations of DCM, and can be extended to fMRI. In addition, future work will demonstrate how to deploy this approach in models for oscillatory decomposition of neural spectra (e.g., \cite{medrano2024bsd}) to enable time-frequency analysis. We hope that this work will allow neuroimaging researchers to construct more advanced models of synaptic dynamics, contributing to a wider effort to uncover fundamental neural mechanisms such as precision-coding and adaption.

\section*{Data and Code Availability}
    The experimental data were published by Mellor et al.~\cite{mellor2023real} and is  \href{https://zenodo.org/records/7872660}{openly available online}. The code associated to the described method is open-source available in SPM at \href{https://github.com/spm/spm}{https://github.com/spm/spm}, with examples provided in files \texttt{DEMO\_tvec\_erp\_sim.m},  \texttt{DEMO\_tvec\_csd\_sim.m}, and  \texttt{DEMO\_tvec\_erp\_mmn.m}. 
    
\section*{Funding}
JM is supported by the Discovery Research Platform for Naturalistic Neuroimaging funded by the Wellcome [226793/Z/22/Z]. PZ is funded by an MRC Career Development Award [MR/X020274/1].

\section*{Acknowledgements}
    We would like to thank Nicholas Alexander for helpful discussions and advice related to the OPM MEG data analysis in this paper.    
\bibliography{references}

\begin{thebibliography}{}

\bibitem[Abraham and Bear, 1996]{abraham1996metaplasticity}
Abraham, W.~C. and Bear, M.~F. (1996).
\newblock Metaplasticity: the plasticity of synaptic plasticity.
\newblock {\em Trends in neurosciences}, 19(4):126--130.

\bibitem[Adam et~al., 2023]{adam2023modulatory}
Adam, E., Kwon, O., Montejo, K.~A., and Brown, E.~N. (2023).
\newblock Modulatory dynamics mark the transition between anesthetic states of unconsciousness.
\newblock {\em Proceedings of the National Academy of Sciences}, 120(30):e2300058120.

\bibitem[Auksztulewicz and Friston, 2016]{auksztulewicz2016repetition}
Auksztulewicz, R. and Friston, K. (2016).
\newblock Repetition suppression and its contextual determinants in predictive coding.
\newblock {\em cortex}, 80:125--140.

\bibitem[Bassett et~al., 2011]{bassett2011dynamic}
Bassett, D.~S., Wymbs, N.~F., Porter, M.~A., Mucha, P.~J., Carlson, J.~M., and Grafton, S.~T. (2011).
\newblock Dynamic reconfiguration of human brain networks during learning.
\newblock {\em Proceedings of the National Academy of Sciences}, 108(18):7641--7646.

\bibitem[Bastos et~al., 2015]{bastos2015dcm}
Bastos, A.~M., Litvak, V., Moran, R., Bosman, C.~A., Fries, P., and Friston, K.~J. (2015).
\newblock A dcm study of spectral asymmetries in feedforward and feedback connections between visual areas v1 and v4 in the monkey.
\newblock {\em Neuroimage}, 108:460--475.

\bibitem[Bastos et~al., 2012]{bastos2012canonical}
Bastos, A.~M., Usrey, W.~M., Adams, R.~A., Mangun, G.~R., Fries, P., and Friston, K.~J. (2012).
\newblock Canonical microcircuits for predictive coding.
\newblock {\em Neuron}, 76(4):695--711.

\bibitem[Breakspear et~al., 2006]{breakspear2006unifying}
Breakspear, M., Roberts, J.~A., Terry, J.~R., Rodrigues, S., Mahant, N., and Robinson, P.~A. (2006).
\newblock A unifying explanation of primary generalized seizures through nonlinear brain modeling and bifurcation analysis.
\newblock {\em Cerebral Cortex}, 16(9):1296--1313.

\bibitem[Brown and Friston, 2013]{brown2013functional}
Brown, H.~R. and Friston, K.~J. (2013).
\newblock The functional anatomy of attention: a dcm study.
\newblock {\em Frontiers in human neuroscience}, 7:784.

\bibitem[Carr, 1981]{carr2012applications}
Carr, J. (1981).
\newblock {\em Applications of centre manifold theory}, volume~35.
\newblock Springer Science \& Business Media.

\bibitem[Ching and Brown, 2014]{ching2014modeling}
Ching, S. and Brown, E.~N. (2014).
\newblock Modeling the dynamical effects of anesthesia on brain circuits.
\newblock {\em Current opinion in neurobiology}, 25:116--122.

\bibitem[Citri and Malenka, 2008]{citri2008synaptic}
Citri, A. and Malenka, R.~C. (2008).
\newblock Synaptic plasticity: multiple forms, functions, and mechanisms.
\newblock {\em Neuropsychopharmacology}, 33(1):18--41.

\bibitem[David et~al., 2005]{david2005modelling}
David, O., Harrison, L., and Friston, K.~J. (2005).
\newblock Modelling event-related responses in the brain.
\newblock {\em NeuroImage}, 25(3):756--770.

\bibitem[Fowlie, 2023]{fowlie2023neyman}
Fowlie, A. (2023).
\newblock Neyman--pearson lemma for bayes factors.
\newblock {\em Communications in Statistics-Theory and Methods}, 52(15):5379--5386.

\bibitem[Friston, 2008]{friston2008hierarchical}
Friston, K. (2008).
\newblock Hierarchical models in the brain.
\newblock {\em PLoS computational biology}, 4(11):e1000211.

\bibitem[Friston et~al., 2012a]{friston2012dcm}
Friston, K.~J., Bastos, A., Litvak, V., Stephan, K.~E., Fries, P., and Moran, R.~J. (2012a).
\newblock Dcm for complex-valued data: cross-spectra, coherence and phase-delays.
\newblock {\em Neuroimage}, 59(1):439--455.

\bibitem[Friston et~al., 2003]{friston2003dynamic}
Friston, K.~J., Harrison, L., and Penny, W. (2003).
\newblock Dynamic causal modelling.
\newblock {\em Neuroimage}, 19(4):1273--1302.

\bibitem[Friston et~al., 2016]{friston2016bayesian}
Friston, K.~J., Litvak, V., Oswal, A., Razi, A., Stephan, K.~E., Van~Wijk, B.~C., Ziegler, G., and Zeidman, P. (2016).
\newblock Bayesian model reduction and empirical bayes for group (dcm) studies.
\newblock {\em Neuroimage}, 128:413--431.

\bibitem[Friston et~al., 2012b]{friston2012dopamine}
Friston, K.~J., Shiner, T., FitzGerald, T., Galea, J.~M., Adams, R., Brown, H., Dolan, R.~J., Moran, R., Stephan, K.~E., and Bestmann, S. (2012b).
\newblock Dopamine, affordance and active inference.
\newblock {\em PLoS computational biology}, 8(1):e1002327.

\bibitem[Garrido et~al., 2008]{garrido2008functional}
Garrido, M.~I., Friston, K.~J., Kiebel, S.~J., Stephan, K.~E., Baldeweg, T., and Kilner, J.~M. (2008).
\newblock The functional anatomy of the mmn: a dcm study of the roving paradigm.
\newblock {\em Neuroimage}, 42(2):936--944.

\bibitem[Haken, 1977]{haken1977synergetics}
Haken, H. (1977).
\newblock Synergetics.
\newblock {\em Physics Bulletin}, 28(9):412.

\bibitem[Haken, 2006]{haken2006synergetics}
Haken, H. (2006).
\newblock Synergetics of brain function.
\newblock {\em International journal of psychophysiology}, 60(2):110--124.

\bibitem[Hutchison et~al., 2013]{hutchison2013dynamic}
Hutchison, R.~M., Womelsdorf, T., Allen, E.~A., Bandettini, P.~A., Calhoun, V.~D., Corbetta, M., Della~Penna, S., Duyn, J.~H., Glover, G.~H., Gonzalez-Castillo, J., et~al. (2013).
\newblock Dynamic functional connectivity: promise, issues, and interpretations.
\newblock {\em Neuroimage}, 80:360--378.

\bibitem[Jafarian et~al., 2021]{jafarian2021adiabatic}
Jafarian, A., Zeidman, P., Wykes, R.~C., Walker, M., and Friston, K.~J. (2021).
\newblock Adiabatic dynamic causal modelling.
\newblock {\em NeuroImage}, 238:118243.

\bibitem[Jansen and Rit, 1995]{jansen1995electroencephalogram}
Jansen, B.~H. and Rit, V.~G. (1995).
\newblock Electroencephalogram and visual evoked potential generation in a mathematical model of coupled cortical columns.
\newblock {\em Biological cybernetics}, 73(4):357--366.

\bibitem[Jefferys, 1995]{jefferys1995nonsynaptic}
Jefferys, J. (1995).
\newblock Nonsynaptic modulation of neuronal activity in the brain: electric currents and extracellular ions.
\newblock {\em Physiological reviews}, 75(4):689--723.

\bibitem[Jirsa et~al., 2014]{jirsa2014nature}
Jirsa, V.~K., Stacey, W.~C., Quilichini, P.~P., Ivanov, A.~I., and Bernard, C. (2014).
\newblock On the nature of seizure dynamics.
\newblock {\em Brain}, 137(8):2210--2230.

\bibitem[Kiebel et~al., 2008]{kiebel2008dynamic}
Kiebel, S.~J., Garrido, M.~I., Moran, R.~J., and Friston, K.~J. (2008).
\newblock Dynamic causal modelling for eeg and meg.
\newblock {\em Cognitive neurodynamics}, 2:121--136.

\bibitem[Lomb, 1976]{lomb1976least}
Lomb, N.~R. (1976).
\newblock Least-squares frequency analysis of unequally spaced data.
\newblock {\em Astrophysics and space science}, 39:447--462.

\bibitem[Magee and Grienberger, 2020]{magee2020synaptic}
Magee, J.~C. and Grienberger, C. (2020).
\newblock Synaptic plasticity forms and functions.
\newblock {\em Annual review of neuroscience}, 43(1):95--117.

\bibitem[Medrano et~al., 2024a]{medrano2024bsd}
Medrano, J., Alexander, N.~A., Seymour, R.~A., and Zeidman, P. (2024a).
\newblock Bsd: a bayesian framework for parametric models of neural spectra.
\newblock {\em arXiv preprint arXiv:2410.20896}.

\bibitem[Medrano et~al., 2024b]{medrano2024linking}
Medrano, J., Friston, K., and Zeidman, P. (2024b).
\newblock Linking fast and slow: The case for generative models.
\newblock {\em Network Neuroscience}, 8(1):24--43.

\bibitem[Mellor et~al., 2023]{mellor2023real}
Mellor, S., Tierney, T.~M., Seymour, R.~A., Timms, R.~C., O'Neill, G.~C., Alexander, N., Spedden, M.~E., Payne, H., and Barnes, G.~R. (2023).
\newblock Real-time, model-based magnetic field correction for moving, wearable meg.
\newblock {\em NeuroImage}, 278:120252.

\bibitem[Moran et~al., 2013a]{moran2013neural}
Moran, R., Pinotsis, D.~A., and Friston, K. (2013a).
\newblock Neural masses and fields in dynamic causal modeling.
\newblock {\em Frontiers in computational neuroscience}, 7:57.

\bibitem[Moran et~al., 2013b]{moran2013free}
Moran, R.~J., Campo, P., Symmonds, M., Stephan, K.~E., Dolan, R.~J., and Friston, K.~J. (2013b).
\newblock Free energy, precision and learning: the role of cholinergic neuromodulation.
\newblock {\em Journal of Neuroscience}, 33(19):8227--8236.

\bibitem[Moran et~al., 2011]{moran2011consistent}
Moran, R.~J., Stephan, K.~E., Dolan, R.~J., and Friston, K.~J. (2011).
\newblock Consistent spectral predictors for dynamic causal models of steady-state responses.
\newblock {\em Neuroimage}, 55(4):1694--1708.

\bibitem[Moran et~al., 2009]{moran2009dynamic}
Moran, R.~J., Stephan, K.~E., Seidenbecher, T., Pape, H.-C., Dolan, R.~J., and Friston, K.~J. (2009).
\newblock Dynamic causal models of steady-state responses.
\newblock {\em Neuroimage}, 44(3):796--811.

\bibitem[Neyman and Pearson, 1933]{neyman1933ix}
Neyman, J. and Pearson, E.~S. (1933).
\newblock Ix. on the problem of the most efficient tests of statistical hypotheses.
\newblock {\em Philosophical Transactions of the Royal Society of London. Series A, Containing Papers of a Mathematical or Physical Character}, 231(694-706):289--337.

\bibitem[Papadopoulou et~al., 2015]{papadopoulou2015tracking}
Papadopoulou, M., Leite, M., van Mierlo, P., Vonck, K., Lemieux, L., Friston, K., and Marinazzo, D. (2015).
\newblock Tracking slow modulations in synaptic gain using dynamic causal modelling: validation in epilepsy.
\newblock {\em Neuroimage}, 107:117--126.

\bibitem[Rosch et~al., 2018]{rosch2018calcium}
Rosch, R.~E., Hunter, P.~R., Baldeweg, T., Friston, K.~J., and Meyer, M.~P. (2018).
\newblock Calcium imaging and dynamic causal modelling reveal brain-wide changes in effective connectivity and synaptic dynamics during epileptic seizures.
\newblock {\em PLoS computational biology}, 14(8):e1006375.

\bibitem[Scargle, 1982]{scargle1982studies}
Scargle, J.~D. (1982).
\newblock Studies in astronomical time series analysis. ii-statistical aspects of spectral analysis of unevenly spaced data.
\newblock {\em Astrophysical Journal, Part 1, vol. 263, Dec. 15, 1982, p. 835-853.}, 263:835--853.

\bibitem[Seymour et~al., 2021]{seymour2021using}
Seymour, R.~A., Alexander, N., Mellor, S., O'Neill, G.~C., Tierney, T.~M., Barnes, G.~R., and Maguire, E.~A. (2021).
\newblock Using opms to measure neural activity in standing, mobile participants.
\newblock {\em NeuroImage}, 244:118604.

\bibitem[Soplata et~al., 2023]{soplata2023rapid}
Soplata, A.~E., Adam, E., Brown, E.~N., Purdon, P.~L., McCarthy, M.~M., and Kopell, N. (2023).
\newblock Rapid thalamocortical network switching mediated by cortical synchronization underlies propofol-induced eeg signatures: a biophysical model.
\newblock {\em Journal of Neurophysiology}, 130(1):86--103.

\bibitem[VanderPlas, 2018]{vanderplas2018understanding}
VanderPlas, J.~T. (2018).
\newblock Understanding the lomb--scargle periodogram.
\newblock {\em The Astrophysical Journal Supplement Series}, 236(1):16.

\bibitem[Vidaurre et~al., 2018]{vidaurre2018discovering}
Vidaurre, D., Abeysuriya, R., Becker, R., Quinn, A.~J., Alfaro-Almagro, F., Smith, S.~M., and Woolrich, M.~W. (2018).
\newblock Discovering dynamic brain networks from big data in rest and task.
\newblock {\em NeuroImage}, 180:646--656.

\bibitem[Wendling et~al., 2016]{wendling2016computational}
Wendling, F., Benquet, P., Bartolomei, F., and Jirsa, V. (2016).
\newblock Computational models of epileptiform activity.
\newblock {\em Journal of neuroscience methods}, 260:233--251.

\bibitem[Zarghami and Friston, 2020]{zarghami2020dynamic}
Zarghami, T.~S. and Friston, K.~J. (2020).
\newblock Dynamic effective connectivity.
\newblock {\em Neuroimage}, 207:116453.

\bibitem[Zeidman et~al., 2019]{zeidman2019guide}
Zeidman, P., Jafarian, A., Seghier, M.~L., Litvak, V., Cagnan, H., Price, C.~J., and Friston, K.~J. (2019).
\newblock A guide to group effective connectivity analysis, part 2: Second level analysis with peb.
\newblock {\em Neuroimage}, 200:12--25.

\end{thebibliography}

\end{document}